\documentstyle[prd,aps,eqsecnum]{revtex}
\newcommand{\bmit}[1]{\mbox{\boldmath $#1$}}

\newcommand{\f}{\varphi}
\newcommand{\F}{\phi}
\newcommand{\zt}{x^3 - x^0}
\newcommand{\mt}{- x^0}
\begin{document}

\draft
\title{Exact plane gravitational waves and electromagnetic fields}
\author{Enrico Montanari \footnote{e--mail address: 
montanari@fe.infn.it.}
and Mirco Calura}
\address{Department of Physics, University of Ferrara and
INFN Sezione di Ferrara, Via Paradiso 12,
I-44100 Ferrara, Italy}
\maketitle
\begin{abstract}
The behaviour of a ``test'' electromagnetic field in the background 
of an exact gravitational plane wave is investigated in the framework 
of Einstein's general relativity. We have expressed the general 
solution to the de Rham equations as a Fourier--like 
integral. In the general case we have reduced the problem to a
set of ordinary differential equations and have explicitly written 
the solution in the case of linear polarization of the gravitational 
wave. 
We have expressed our results by means of Fermi Normal Coordinates 
(FNC), which 
define the proper reference frame of the laboratory. Moreover we have 
provided some {\em gedanken experiments}, showing that an external 
gravitational wave induces
measurable effects of non tidal nature
via electromagnetic interaction. Consequently it is 
not possible to eliminate gravitational effects
on electromagnetic field, even in an 
arbitrarily small spatial region around an observer freely falling in 
the field of a gravitational wave. This is opposite to the case of 
mechanical 
interaction involving measurements of geodesic deviation effects. 
This behaviour is not in contrast with the principle of 
equivalence, which applies to arbitrarily small region of both 
space and time.
\end{abstract}

\section{Introduction}

The behaviour of a ``test'' electromagnetic field ({\em i.e.} an 
electromagnetic field whose stress--energy tensor does not affect 
the curvature of the underlying space--time) in a 
gravitational wave background has been widely studied in the framework 
of Einstein's general relativity. 
This topic is very important to conceive possible further 
experimental verification of general relativity and also to better
understand the principles underlying the theory itself.
A great deal of efforts have been aimed at solving  
Maxwell (or de Rham) equations to first order in the gravitational 
wave amplitude~\cite{coo68,boc70,
ber71,bra73,fun78,cod80,bar85,bra90,lob92,coo93,mon97,cal98}
or, at most, to second order under geometrical 
optics limit~\cite{zip66}.
However, a general solution within the framework 
of the full theory of general relativity  would  
highlight the main features of a free electromagnetic field in 
radiative curved space--time. Indeed any possible 
ambiguity arising from approximation procedures could be 
circumvented~\cite{mon97,cal98}.
We point out that such ambiguities exist not only when strong 
gravitational waves are concerned, but also may appear for weak
gravitational fields as well. An example of this is the question
related to photon creation [see discussion in Sec.~\ref{sec6} after
Eq.~(\ref{poin})].
Therefore any result obtained in the framework of linearized general 
relativistic theory has to be handle with care due to the non linear 
character of the full Einstein theory. The linearized theory is not 
general relativity but minkowski flat space--time plus a small 
perturbation. 

To solve the problem one can take advantage of the
freedom in the choice of a particular system of coordinates in which
writing the equations. Obviously it is convenient to choose the 
reference frame in which the equations are easier to be solved.
This is usually accomplished when the metric tensor is expressed 
in its simplest form. This system {\em attached} to the wave will be
hereinafter
referred to as Wave Reference Frame (WRF). In the limit of linearized 
gravity, it reduces to the usual Transverse Traceless (TT) 
gauge~\cite{mtw}.
Recently a solution in the WRF has been obtained~\cite{bal97}. 
However, such a system is not the one where measurements are performed
(laboratory frame).
Therefore, in order to obtain results that have a direct 
interpretation from the physics point of view, one should express the 
electromagnetic
field variables in a reference frame attached to an observer. The most 
natural way to construct such a frame is to consider the observer 
freely falling in the field of the gravitational wave. Such 
coordinates are the well--known Fermi Normal Coordinates 
(FNC)~\cite{mtw,for82}. 

The aim 
of this paper is to obtain the solution of this problem in FNC and 
employ this solution to predict a few new physical effects. To this 
purpose we have re--obtained the solution in WRF in a slightly different
way, expressing it as a Fourier--like integral. This form is more 
suitable for applications to concrete physical situations. Moreover we 
have obtained the transformation rules between WRF and FNC which hold 
true in each point of the region of space--time where the plane 
propagation of the exact gravitational wave is valid.

One of the most important results is that in FNC it is possible to 
infer the presence of a gravitational wave over an arbitrarily small 
spatial 
region in the neighbourhood of the origin. In other words there 
are both tidal and non--tidal effects in spatial coordinates. 
Therefore, near the origin, d'Alembertian operator applied to the 
electromagnetic field is non--vanishing, but rather proportional to 
the order of magnitude 
of the electromagnetic field. This is not surprising because 
Maxwell second--order equations
involve the Riemann tensor~\cite{mtw,tol}. 
Consequently the solution to Maxwell equations can not be  
written as a monochromatic plane wave and no photon is created by the 
interaction. 
We stress that this behaviour is not in contrast with the principle of 
equivalence, which applies to arbitrarily small region of 
space--time~\cite{mtw}. This topic is more deeply investigated 
in~\cite{prl}.

The paper is organized as follows. In Secs. II and III we perform 
calculation in WRF. In Sec. IV we provide the transformation rules 
connecting WRF and FNC. In Sec. V we express the four--vector 
potential in the latter frame. Finally, in Sec. VI 
the theory is applied to a case that is theoretically 
interesting and opens up some {\em gedanken experiments}.

\section{Solution to the free de Rham equations in WRF}

Let us call $y^{\mu}$ the coordinates in WRF.
According to~\cite{mtw,l2,sch} an exact plane gravitational wave 
propagating along the 
$y^3$--axis is described by the following  line 
element
\begin{eqnarray}
ds^2 &=& - (y^0)^2 + (y^3)^2 + 
{\cal P}(y^3-y^0) (dy^1)^2 \label{dsTT} 
\label{element}\\ 
&&+{\cal Q}(y^3-y^0) (dy^2)^2  + 
2\,{\cal S}(y^3-y^0) dy^1 dy^2 \nonumber
\end{eqnarray}
We do not make any assumption about the actual 
form or amplitude of the three quantities ${\cal P}, {\cal Q}$, and ${\cal S}$ 
but simply require that the metric tensor is an exact solution to the vacuum 
Einstein equations.
Moreover, were the curvature small, 
the line element (\ref{element}) would just describe a weak plane 
waves in a {\em TT} reference frame~\cite{mtw}.  
Since metric coefficients depend on the coordinates only through 
$y^3-y^0$ we are naturally led to perform the following transformation:
\begin{eqnarray}
Y^0&=&y^3 - y^0\nonumber\\
Y^1&=&y^1\nonumber\\
Y^2&=&y^2\nonumber\\
Y^3&=&y^3 + y^0
\label{trasformation}
\end{eqnarray}
Since the wave keeps plane in the new set of coordinates,
the new reference frame $Y^{\mu}$ will 
still be referred to as $WRF$. The covariant components of the metric
tensor become: 
\begin{eqnarray}
g_{\mu\nu} = 
\left(
\begin{array}{cccc}
 0 & 0 & 0 & \frac{1}{2}\\
 0 & {\cal P}\left(Y^0\right) & {\cal S}\left(Y^0\right)&0\\
 0 & {\cal S}\left(Y^0\right)& {\cal Q}\left(Y^0\right)&0\\
 \frac{1}{2}&0 & 0& 0
\end{array}
\right)
\label{G}
\end{eqnarray}
In order to have a real space--time, 
the following inequality must hold true:
\begin{equation}
g = -\frac{1}{4}\,\left({\cal P}{\cal Q} - {\cal S}^2
\right) < 0
\label{realspace}
\end{equation}
First of all we consider the general form of de Rham equations
under Lorentz condition, describing the propagation of free 
electromagnetic field in empty curved space--time (we stress again 
that the electromagnetic field strength is considered so
small as to neglect its effects on the curvature of space--time):
\begin{eqnarray}
{\cal A}^{\alpha ;\nu}_{\ \ \ ;\nu} &=&0
\label{derhamgen1}\\
{\cal A}^{\nu}_{\ ;\nu}&=&0
\label{derhamgen2}
\label{derhamgen}
\end{eqnarray}
By substituting the metric tensor~(\ref{G}) 
in~(\ref{derhamgen1})--(\ref{derhamgen2}) we achieve
(we use convention and notation as in~\cite{mtw}
except for $p$, $q$, $r$, and $s$ which take values
1 and 2):
\begin{eqnarray}
&&g^{\mu\nu}{\cal A}^{\alpha}_{\ ,\mu,\nu}+2\,g^{\alpha q}\dot{g}_{qp}
\partial_3 {\cal A}^{p}+
g^{\alpha p}g^{qr}\dot{g}_{pq}\partial_r {\cal A}^0 -
2\,\delta^{\alpha}_{\ 3}\,
g^{qp}\dot{g}_{rq}\partial_p {\cal A}^{r}\nonumber\\
&&-\frac{1}{2}\,\delta^{\alpha}_{\ 3}\,
g^{rs}g^{pq}\dot{g}_{qr}\dot{g}_{ps}\,{\cal A}^0 +
2\,\frac{d\lg{\sqrt{(-g)}}}{dY^0}\,
\partial_3 {\cal A}^{\alpha} = 0,
\label{derham}
\end{eqnarray}
with Lorentz condition:
\begin{equation}
\partial_{\mu}{\cal A}^{\mu}+\frac{d\lg{\sqrt{(-g)}}}{dY^0}\,
{\cal A}^0 = 0;
\label{lorentz}
\end{equation}
dot means derivation with respect to the argument.
As is well known in classical electrodynamics ({\em e.g.}~\cite{jac}),
even for potential satisfying Lorentz condition there remains the 
possibility to perform a restricted gauge transformation such that the 
resulting new potential still satisfies the Lorentz condition. 

Our aim is to solve the system of equations~(\ref{derham}) taking into 
account condition~(\ref{lorentz}); in order to 
simplify them, we choose the particular restricted Lorentz
gauge in which ${\cal A}^0 = 0$. We will show that this choice is 
always possible, without loss of generality (analogously to the 
classical charge--free case~\cite{jac}). In fact, if ${\cal A}'^\mu$
satisfies Eq.~(\ref{derham}) with the Lorentz 
condition~(\ref{lorentz}), then also 
${\cal A}^\mu + \Lambda^{;\mu}$ does, provided that:
\begin{equation}
g^{\alpha\nu}\,\Lambda_{,\nu,\alpha} + 
2\,\frac{d\lg{\sqrt{(-g)}}}{dY^0}\,\Lambda_{,3} = 0,
\label{lambda}
\end{equation}
where $\Lambda$ is a scalar function.
Setting ${\cal A}^0 = 0$, then 
$\Lambda_{,3} = \frac{1}{2}\,{\cal A}'^0$. If this last condition 
holds, then, by substitution in Eq.~(\ref{derham}) 
for $\alpha=0$, it follows that $\Lambda$ indeed does satisfy:
\begin{equation}
\left (
g^{\alpha\nu}\,\Lambda_{,\nu,\alpha} +
2\,\frac{d\lg{\sqrt{(-g)}}}{dY^0}\,\Lambda_{,3} 
\right )_{,3} = 0.
\label{lambda2}
\end{equation}
In conclusion, for the charge--free case, it is always possible to set 
$A^0=0$.

Because of this choice both  
Eqs.~(\ref{derham}) and~(\ref{lorentz}) turn into simpler 
ones. Namely:
\begin{equation}
g^{\mu\nu}{\cal A}^{\alpha}_{\ ,\mu,\nu}+2\,g^{\alpha q}\dot{g}_{qp}
\partial_3 {\cal A}^{p}-
2\,\delta^{\alpha}_{\ 3}\,
g^{qp}\dot{g}_{rq}\partial_p {\cal A}^{r}
+2\,\frac{d\lg{\sqrt{(-g)}}}{dY^0}\,
\partial_3 {\cal A}^{\alpha} = 0
\label{newderham}
\end{equation}  
and:
\begin{equation}
\partial_{k}{\cal A}^{k} = 0
\label{newlorentz}
\end{equation}

The advantage of having replaced the old variables $y^\alpha$ with 
$Y^\alpha$ is that in Eqs.~(\ref{newderham}) there is no
second--order derivative with respect to either $Y^0$ or $Y^3$. This
fact, together with the dependence of the 
coefficients of the metric--tensor components 
only on $Y^0$, make it possible to turn the partial 
derivatives system~(\ref{newderham}) into a simple ordinary one. 
If we write the solution as a Fourier integral
\begin{eqnarray}
{\cal A}^{\alpha} = \int d^{3}\lambda\,
e^{\left[i(\lambda_j Y^j)\right]}\,
a^{\alpha}\left(Y^0, \lambda_1, \lambda_2, \lambda_3\right)
\label{fourierA}
\end{eqnarray}  
and substitute it in the first two equations of 
system~(\ref{newderham}), we get rid of the dependence on $Y^1$, 
$Y^2$ and $Y^3$ and obtain the following ordinary system for  
{\em Fourier--coefficients} $a^1$ and $a^2$:
\begin{eqnarray}
4i\lambda_3\,\frac{da^s}{dY^0} - \left(\lambda_1^{\ 2} + 
\lambda_2^{\ 2}\right)\,a^s + f^{pq}\lambda_p \lambda_q a^s +
2i\lambda_3\,g^{sq}\dot{g}_{pq} a^p +
i\lambda_3 \frac{d\lg{(-g)}}{dY^0}\,a^s = 0
\label{eqar}
\end{eqnarray}
where we have set $f^{qp} = \delta^{qp} - g^{qp}$.
The above system of equations can still be simplified by setting 
\begin{equation}
a^s = \gamma B^s
\label{Bdef}
\end{equation}
with
\begin{equation}
\gamma = \frac{1}{(-g)^{\frac{1}{4}}}\,
\exp{\left[\frac{\left(\lambda_1^{\ 2} +\lambda_2^{\ 2}\right)\,Y^0 -
f^{(-1)pq} \lambda_p \lambda_q}{4i\lambda_3}\right]}
\label{gammadef}
\end{equation}  
where $f^{(-1)pq}$ is any primitive of $f^{pq}$. 
Equations~(\ref{eqar}) yield the following system of equations for the 
unknown quantities $B^s$:
\begin{equation}
\frac{d B^s}{dY^0} + \frac{1}{2}\,{\cal M}^{s}_{\ p}B^p = 0
\label{eqbs}
\end{equation}
where we have set:
\begin{equation}
{\cal M}^{s}_{\ p} = g^{sq}\dot{g}_{qp}
\label{mdef}
\end{equation}
before discussing the solutions of system~(\ref{eqbs}) we 
consider last equation of the 
system~(\ref{newderham}):
\begin{eqnarray}
4i\lambda_3 \frac{da^3}{dY^0} - \left(\lambda_1^{\ 2} +
\lambda_2^{\ 2}\right)\,a^3 + f^{pq}\lambda_p \lambda_q a^3 - 
2i\lambda_p g^{pq}\dot{g}_{qs} a^s +i\lambda_3 \,\frac{d\log{(-g)}}{dY^0}\,
 a^3 = 0
\label{eqa0}
\end{eqnarray}
The solution of this equation comes directly from Lorentz condition: 
\begin{equation}
a^{0} = 0;\ \ a^{3} = \gamma B^3;\qquad B^3 = - \frac{\lambda_r B^r}
{\lambda_3} 
\label{a0}
\end{equation}

Now only Eq.~(\ref{eqbs}) needs to be solved: 
once $B^{1}$ and $B^{2}$ have been obtained, it is trivial to
achieve $A^{\alpha}$ by means of Eqs.~(\ref{Bdef}), (\ref{a0}) 
and~(\ref{fourierA}). The solution to Eq.~(\ref{eqbs}) can be more or 
less difficult to be determined depending on the actual form of 
${\cal P}(Y^0)$, ${\cal Q}(Y^0)$, and ${\cal S}(Y^0)$.
In the general case, an analytical solution 
to Eqs.~(\ref{eqbs}) could not be available. A major advantage of 
this system is its non--dependence on 
$\lambda_{k}$; therefore a numerical solution can be easily 
implemented. The general solution to Eqs.~(\ref{eqbs}) can be inferred 
by multiplying the numerical solution by an arbitrary 
function of $\lambda_k$.

The results of this Section are in agreement with those obtained 
in~\cite{bal97}. Yet, we have used a different formalism which is 
more suitable for application to concrete physical situations. For
instance, by choosing properly the Fourier--like coefficients of 
Eq.~(\ref{fourierA}), any possible particular boundary or initial 
condition can be obtained.

In next Section we will show that it is possible to 
solve exactly such a system for the case of a linearly polarized 
gravitational wave.

\section{Linear polarization}

An exact plane gravitational wave is linearly polarized when 
the metric tensor can be put into diagonal form through a fixed
rotation in the $Y^1$--$Y^2$ plane. 

Consequently, without losing generality, we choose the $Y^\mu$ 
reference frame in such a way that the only non--vanishing components 
of the metric tensor are:
\begin{eqnarray}
&&g_{03} = g_{30} = \frac{1}{2}\nonumber\\
&&g_{11} = {\cal P}(Y^0),\qquad g_{22} = {\cal Q}(Y^0). 
\label{Gdiag}
\end{eqnarray}

If we set ${\cal P}(Y^0) = L^2(Y^0)\,\exp({2\beta(Y^0)})$ and 
${\cal Q}(Y^0) = L^2(Y^0)\,\exp({-2\beta(Y^0)})$, $L(Y^0)$ and 
$\beta(Y^0)$ being arbitrary functions, then the condition for the 
gravitational field equations to be satisfied becomes~\cite{mtw}:
\begin{equation}
\frac{d^2 L}{d(Y^0)^2} + \left(\frac{d\beta}{dY^0}\right)^2\,L = 0
\label{Leq}
\end{equation}
One can easily see that, were $\beta$ small, the metric would reduce 
to that of a usual weak gravitational wave in linearized 
theory~\cite{mtw}. 
According to~\cite{mtw}, $L(Y^0)$ and $\beta(Y^0)$ are referred to as
{\em background} and  {\em wave factors}.

We now proceed to calculate the exact solution to the system of 
equations~(\ref{eqbs}). By using the background and wave factors instead 
of ${\cal P}$ and ${\cal Q}$, it reads:
\begin{eqnarray}
&&\frac{dB^1}{dY^0} + \frac{1}{L}\,\left(
\frac{dL}{dY^0} + L(Y^0)\,\frac{d\beta}{dY^0}\right) B^1(Y^0) = 0\\
\label{eqbs1}
&&\frac{dB^2}{dY^0} + \frac{1}{L}\,\left(
\frac{dL}{dY^0} - L(Y^0)\,\frac{d\beta}{dY^0}\right) B^2(Y^0) = 0.     
\label{eqbs2}
\end{eqnarray}
If we set:
\begin{equation}
{\cal V}^+ = \beta + \frac{1}{2}\,\lg{(L^2)},\qquad\qquad
{\cal V}^- = \beta - \frac{1}{2}\,\lg{(L^2)}
\end{equation}
after some manipulation, we obtain:
\begin{eqnarray}
&&\frac{dB^1}{B^1} = - d{\cal V}^+\\
&&\frac{dB^2}{B^2} =   d{\cal V}^-.
\end{eqnarray}
These equations can be easily integrated obtaining:
\begin{eqnarray}
&&B^1(Y^0) = b^1\,e^{-{\cal V}^+}\label{solb1}\\
&&B^2(Y^0) = b^2\,e^{{\cal V}^-}
\label{solb2}
\end{eqnarray}
where $b^1$ and $b^2$ are constants. Substitution of 
Eqs.~(\ref{solb1}) and~(\ref{solb2}) in Eqs.~(\ref{a0}) yields:
\begin{equation}
B^3 = - \frac{\lambda_1}{\lambda_3}\,b^1\,e^{-{\cal V}^+} - 
\frac{\lambda_2}{\lambda_3}\,b^2\,e^{{\cal V}^-}.\label{solb3}
\end{equation}
As for Eq.~(\ref{gammadef}), we obtain
\begin{equation}
\gamma = \sqrt{2}\,e^{-\frac{1}{2}({\cal V}^+ - {\cal V}^-) + 
\frac{1}{4i\lambda_3}\,\left((\lambda_1)^2 V^+ 
 + (\lambda_2)^2 V^-\right)}\label{gammadiag}
\end{equation}
where $V^{+}$ and $V^{-}$ are functions defined as follows:
\begin{equation}
\frac{dV^{\pm}(w)}{dw} = e^{\mp 2{\cal V}^{\pm}(w)}
\end{equation}

\section{FNC for an exact linearly polarized wave}

The results we have obtained in last section are valid in the WRF; 
though they do not correspond to any measurable quantity. 
In fact, the proper reference frame where an observer may execute a 
measurement is the FNC~\cite{mtw,for82}.
Therefore any tensor expressed in 
WRF must be written in FNC. To this aim, this section is devoted to 
linking the WRF to the laboratory reference frame (FNC); 
this is accomplished by following the procedure outlined 
in~\cite{mtw,for82}.
First, we consider an observer moving along a time--like geodesic in
WRF; let $q(\tau)$ be its world line, $\tau$ its proper time, and 
$f_{(0)}$ its four--velocity. Thus we can write: 
\begin{equation}
f_{(0)}^{\alpha} = \frac{dq^\alpha(\tau)}{d\tau}.
\end{equation}
To achieve the actual form of the geodesic the following 
equations need to be solved:
\begin{eqnarray}
\frac{df_{(0)}^{\alpha}}{d\tau} + 
\Gamma^{\alpha}_{\mu\nu}f_{(0)}^{\mu}f_{(0)}^{\nu} &=& 0\label{tlg}\\
g_{\mu\nu} f_{(0)}^{\mu} f_{(0)}^{\nu} &=& -1 \label{00}
\end{eqnarray}
The only non--vanishing Christoffel symbols are:
\begin{eqnarray}
&&\Gamma^{1}_{01} = \frac{\partial {\cal V}^+}{\partial Y^0}\qquad\qquad
\Gamma^{2}_{02} = -\frac{\partial {\cal V}^-}{\partial Y^0}\nonumber\\
&&\Gamma^{3}_{11} = -\frac{\partial}{\partial Y^0}\left(
e^{2{\cal V}^{+}(Y^0)}\right)\qquad\qquad
\Gamma^{3}_{22} = -\frac{\partial}{\partial Y^0}\left(
e^{-2{\cal V}^{-}(Y^0)}\right) 
\end{eqnarray}
After straightforward calculation we obtain:
\begin{eqnarray}
f_{(0)}^{0}(\tau)&=&\f_{(0)}^0\\
f_{(0)}^{1}(\tau)&=&\f_{(0)}^1 \, e^{-2 {\cal V}^+ (\f_{(0)}^0\,\tau)}
\\
f_{(0)}^{2}(\tau)&=&\f_{(0)}^2 \, e^{2 {\cal V}^- (\f_{(0)}^0\,\tau)}
\\ 
f_{(0)}^{3}(\tau)&=&-\frac{(\f_{(0)}^1)^2}{\f_{(0)}^0}\,
e^{-2 {\cal V}^+ (\f_{(0)}^0\,\tau)} - 
\frac{(\f_{(0)}^2)^2}{\f_{(0)}^0}\,
e^{2 {\cal V}^- (\f_{(0)}^0\,\tau)} + \f_{(0)}^3,
\end{eqnarray}
where $\f_{(0)}^\alpha$ are four constants.
The geodesic parameterization may be achieved by performing the 
following integration:
\begin{equation}
q^\alpha(\tau) = q^\alpha(0) + \int^\tau_0{f^\alpha_{(0)}(\tau')\,
d\tau'}.
\label{qalpha}
\end{equation}

Once $f_{(0)}$ has been determined, which is the time--like vector of the 
orthonormal tetrad carried by the observer, we  
determine the other three orthonormal space--like vectors $f_{(i)}$, 
to complete the observer's orthonormal tetrad~\cite{mtw,for82}.
These vectors have to be parallel transported along $q(\tau)$. Therefore 
the following set of equations must hold true:
\begin{equation}
\frac{df_{(i)}^{\alpha}}{d\tau} +
\Gamma^{\alpha}_{\mu\nu}f_{(i)}^{\mu}f_{(0)}^{\nu} = 0.
\end{equation}
The solution can be written as:
\begin{eqnarray}
f_{(i)}^{0}(\tau)&=&\f_{(i)}^0\\
f_{(i)}^{1}(\tau)&=&\frac{\f_{(i)}^0 \f_{(0)}^1}{\f_{(0)}^0}\,
e^{-2 {\cal V}^+ (\f_{(0)}^0\,\tau)} + 
\f_{(i)}^1 \,e^{-{\cal V}^+ (\f_{(0)}^0 \,\tau)}\\
f_{(i)}^{2}(\tau)&=&\frac{\f_{(i)}^0 \f_{(0)}^2}{\f_{(0)}^0}\,
e^{2 {\cal V}^- (\f_{(0)}^0\,\tau)} +
\f_{(i)}^2 \,e^{{\cal V}^- (\f_{(0)}^0 \,\tau)}\\
f_{(i)}^{3}(\tau)&=&-\frac{\f_{(i)}^0 (\f_{(0)}^1)^2}{(\f_{(0)}^0)^2}
 \,e^{-2 {\cal V}^+ (\f_{(0)}^0\,\tau)}
-\frac{\f_{(i)}^0 (\f_{(0)}^2)^2}{(\f_{(0)}^0)^2}
 \,e^{2 {\cal V}^- (\f_{(0)}^0\,\tau)}\nonumber\\ 
&-&2\,\frac{\f_{(i)}^1 \f_{(0)}^1}{\f_{(0)}^0} 
\,e^{-{\cal V}^+ (\f_{(0)}^0\,\tau)} - 2\,\frac{\f_{(i)}^2 \f_{(0)}^2}{\f_{(0)}^0}
\,e^{{\cal V}^- (\f_{(0)}^0\,\tau)} + \f_{(i)}^3
\end{eqnarray}
where $\f_{(j)}^\alpha$ are twelve constants.
Besides, orthonormality conditions imply:
\begin{equation}
g_{\mu\nu} f_{(0)}^{\mu} f_{(i)}^{\nu} = 0\qquad\qquad
g_{\mu\nu} f_{(i)}^{\mu} f_{(j)}^{\nu} = \delta_{ij}.
\label{0jk}
\end{equation}
The conditions given in Eqs.~(\ref{00}) and~(\ref{0jk}) hold true 
provided that:
\begin{eqnarray}
&&\f_{(i)}^3 \,\f_{(0)}^0 + \f_{(i)}^0 \,\f_{(0)}^3 = 0,\nonumber\\
&&\f_{(i)}^0 \,\f_{(j)}^3 + \f_{(i)}^3 \,\f_{(j)}^0 +
2\,\f_{(i)}^1 \,\f_{(j)}^1 + 2\,\f_{(i)}^2 \,\f_{(j)}^2 = 
2\,\delta_{ij},\label{condfi}\\
&&\f_{(0)}^0 \,\f_{(0)}^3 = -1.\nonumber
\end{eqnarray}  

Next step is construction of a space--like geodesic, 
originating from an arbitrary point on the observer world line.
Let $\sigma$ be the space--like geodesic parameter; therefore
$Y(\sigma,\tau)$ is the point parameterized by the 
$\sigma$--variable on the 
space--like geodesic whose emanation point is $q(\tau)$. 
Let $f^{\alpha} = \frac{dY^{\alpha}}{d\sigma}$ be the tangent vector 
in $Y(\sigma,\tau)$ to the space--like geodesic; the geodesic equation 
reads:
\begin{equation}
\frac{df^{\alpha}}{d\sigma} +
\Gamma^{\alpha}_{\mu\nu}f^{\mu}f^{\nu} = 0.
\end{equation}
We seek out the solution to the above equation satisfying the following 
conditions at $\sigma=0$:
\begin{eqnarray}
f^{\mu}(\sigma = 0)&=&\alpha^{j}\,f_{(j)}^{\mu}(\tau),\\
Y^{\mu}(\sigma = 0)&=&q^{\mu}(\tau),
\end{eqnarray}
where $\alpha^{j}$ are the direction cosines of the geodesic at 
$q(\tau)$. 
In order to achieve $f$ we perform
a calculation, which is formally very similar to the one we 
made to solve Eq.~(\ref{tlg});
then we obtain $Y^{\alpha}(\sigma,\,\tau)$ by direct integration: 
\begin{equation}
Y^{\alpha}(\sigma,\,\tau) = q^{\alpha}(\tau) + 
\int_{0}^{\sigma}\,f^{\alpha}\,d\sigma'
\label{Yst}
\end{equation}

A natural way to identify an event in the 
observer's reference frame is taking the four numbers~\cite{mtw}:
\begin{equation}
x^0 = \tau \qquad\qquad x^k = \alpha^k \,\sigma
\end{equation} 
Substitution of these identities in Eq.~(\ref{Yst}) yields the desired 
coordinate transformation:
\begin{eqnarray}
Y^0&=&\f_{(0)}^0 x^0 + \f_{(k)}^0 x^k\label{tt0fnc}\\
Y^1&=&\frac{(\f_{(k)}^1 x^k)}{(\f_{(k)}^0 x^k)}\,
e^{{\cal V}^+ (\f_{(0)}^0 x^0)}\,\left[
V^+ (\f_{(0)}^0 x^0 + \f_{(k)}^0 x^k) - 
V^+ (\f_{(0)}^0 x^0)\right] \nonumber\\
&&+\frac{\f_{(0)}^1}{\f_{(0)}^0}\,\left[
V^+ (\f_{(0)}^0 x^0 + \f_{(k)}^0 x^k) -
V^+ (0)\right] + q^1 (0)\label{tt1fnc}\\
Y^2&=&\frac{(\f_{(k)}^2 x^k)}{(\f_{(k)}^0 x^k)}\,
e^{-{\cal V}^- (\f_{(0)}^0 x^0)}\,\left[
V^- (\f_{(0)}^0 x^0 + \f_{(k)}^0 x^k) -
V^- (\f_{(0)}^0 x^0)\right] \nonumber\\
&&+\frac{\f_{(0)}^2}{\f_{(0)}^0}\,\left[
V^- (\f_{(0)}^0 x^0 + \f_{(k)}^0 x^k) -
V^- (0)\right] + q^2 (0)\label{tt2fnc}\\
Y^3&=&-\frac{(\f_{(0)}^1)^2}{(\f_{(0)}^0)^2}\,
\left[V^+ (\f_{(0)}^0 x^0 + \f_{(k)}^0 x^k) -
V^+ (0)\right] - \frac{(\f_{(0)}^2)^2}{(\f_{(0)}^0)^2}\,
\left[V^- (\f_{(0)}^0 x^0 + \f_{(k)}^0 x^k) -
V^- (0)\right]\nonumber\\ 
&-&\left[\frac{(\f_{(k)}^1 x^k)^2}{(\f_{(k)}^0 x^k)^2}\,
e^{2{\cal V}^+ (\f_{(0)}^0 x^0)} + 
2\,\frac{\f_{(k)}^1 x^k}{\f_{(k)}^0 x^k}\,
\frac{\f_{(0)}^1}{\f_{(0)}^0}\,
e^{{\cal V}^+ (\f_{(0)}^0 x^0)}\right]\,
\left[V^+ (\f_{(0)}^0 x^0 + \f_{(k)}^0 x^k) -
V^+ (\f_{(0)}^0 x^0)\right]\nonumber\\ 
&-&\left[\frac{(\f_{(k)}^2 x^k)^2}{(\f_{(k)}^0 x^k)^2}\,
e^{-2{\cal V}^- (\f_{(0)}^0 x^0)} +
2\,\frac{\f_{(k)}^2 x^k}{\f_{(k)}^0 x^k}\,
\frac{\f_{(0)}^2}{\f_{(0)}^0}\,
e^{-{\cal V}^- (\f_{(0)}^0 x^0)}\right]\,
\left[V^- (\f_{(0)}^0 x^0 + \f_{(k)}^0 x^k) -
V^- (\f_{(0)}^0 x^0)\right]\nonumber\\ 
&+&\f_{(k)}^3 x^k + 
\frac{(\f_{(k)}^1 x^k)^2}{\f_{(k)}^0 x^k} +
\frac{(\f_{(k)}^2 x^k)^2}{\f_{(k)}^0 x^k} +
\f_{(0)}^3 x^0 + q^3 (0) \label{tt3fnc}
\end{eqnarray} 
In the framework of weak--gravitational--field approximation
a similar result was found in~\cite{for82}.

The metric tensor components in FNC could be written through
the usual relation
\begin{equation}
\gamma_{\mu\nu} = \frac{\partial Y^\alpha}{\partial x^\mu}\,
\frac{\partial Y^\beta}{\partial x^\nu}\,g_{\alpha\beta}.
\label{gfnc}
\end{equation}
Yet, for our purposes, it is sufficient to know that 
$\gamma_{\mu\nu}$
$=\eta_{\mu\nu} + {\cal O}\left(|\bmit x|^2\right)$ and we do not 
need to calculate them explicitly.
It can also be checked that
$\Gamma^\mu_{\alpha\beta} = {\cal O}\left(|\bmit x|\right)$.

Although coordinate transformation~(\ref{tt0fnc})--(\ref{tt3fnc})
was determined in the case of a diagonal metric tensor in WRF, 
the expression we obtained holds true 
for a generic linearly polarized 
plane wave.
Consequently Eq.~(\ref{Gdiag}) 
does not lead to any loss in generality.

\section{Four--vector potential in FNC}

The knowledge of the transformation rules given by 
Eqs.~(\ref{tt0fnc})--(\ref{tt3fnc}) allows one to obtain --- 
via direct differentiation --- the expressions of 
$\frac{\partial Y^\alpha}{\partial x^\mu}$ needed to calculate
covariant components of a generic tensor in the FNC. 
Therefore the covariant components of the four--vector potential 
in the WRF are to be determined.
From Eqs.~(\ref{G}),
(\ref{fourierA}), (\ref{Bdef}), (\ref{a0}), 
(\ref{solb1})--(\ref{gammadiag}) we get:
\begin{eqnarray}
a_0&=&-\frac{1}{\sqrt{2}\,\lambda_3}\,\left(
\lambda_1 b^1 e^{-{\cal V}^+ (Y^0)} +
\lambda_2 b^2 e^{{\cal V}^- (Y^0)}\right)\,
e^{-\frac{1}{2}\,({\cal V}^+ (Y^0) - {\cal V}^- (Y^0))}\,
e^{\frac{(\lambda_1)^2 {\cal V}^+ (Y^0) + 
(\lambda_2)^2 {\cal V}^- (Y^0)}{4i\lambda_3}}\\
a_1&=&\sqrt{2}\,b^1 
\,e^{\frac{1}{2}\,({\cal V}^+ (Y^0) + {\cal V}^- (Y^0))}\,
e^{\frac{(\lambda_1)^2 {\cal V}^+ (Y^0) +
(\lambda_2)^2 {\cal V}^- (Y^0)}{4i\lambda_3}}\\
a_2&=&\sqrt{2}\,b^2
\,e^{-\frac{1}{2}\,({\cal V}^+ (Y^0) + {\cal V}^- (Y^0))}\,
e^{\frac{(\lambda_1)^2 {\cal V}^+ (Y^0) +
(\lambda_2)^2 {\cal V}^- (Y^0)}{4i\lambda_3}}\\
a_3&=&0
\end{eqnarray}

By setting $A_\mu$ as the covariant components of the four--vector 
potential in the FNC, they can be achieved by:
\begin{equation}
A_{\mu} = \frac{\partial Y^\alpha}{\partial x^\mu}\,{\cal A}_\alpha.
\end{equation}
The components are written as:
\begin{equation}
A_{\mu}=\int d^{3}\lambda \,e^{i\Psi}\,\bar{a}_{\mu},
\label{fourierfnc}
\end{equation}
where
\begin{eqnarray}
\bar{a}_{0}&=&\left[- \frac{\f_{(0)}^0 \lambda_1}
{\sqrt{2}\lambda_3} + \sqrt{2} \f_{(0)}^0 \,
\frac{\f_{(k)}^1 x^k}{\f_{(k)}^0 x^k}\,e^{{\cal V}^+ (\F_0)} + 
\sqrt{2} \f_{(0)}^1 \right]\,b^1 \, 
e^{-\frac{1}{2}\left(3{\cal V}^+ (\F) -
{\cal V}^- (\F)\right)} \label{afnc0}\\
&+&\left[- \frac{\f_{(0)}^0 \lambda_2}
{\sqrt{2}\lambda_3} + \sqrt{2} \f_{(0)}^0 \,
\frac{\f_{(k)}^2 x^k}{\f_{(k)}^0 x^k}\,e^{-{\cal V}^- (\F_0)} +
\sqrt{2} \f_{(0)}^2 \right]\,b^2 \,
e^{-\frac{1}{2}\left({\cal V}^+ (\F) -
3 {\cal V}^- (\F)\right)} \nonumber \\
&+&\sqrt{2} \f_{(0)}^0\,
\left[\dot{\cal V}^+(\F_0) 
e^{{\cal V}^+ (\F_0)}\,\left(V^+ 
(\F) - V^+ (\F_0)\right) - 
e^{- {\cal V}^+ (\F_0)}\right]\frac{\f_{(k)}^1 x^k}{\f_{(k)}^0 x^k}\,b^1 \,
e^{\frac{1}{2}\left({\cal V}^+ (\F) +
{\cal V}^- (\F)\right)}\nonumber \\
&-&\sqrt{2} \f_{(0)}^0\,
\left[\dot{\cal V}^- (\F_0)
e^{-{\cal V}^- (\F_0)}\,\left(V^-
(\F) - V^- (\F_0)\right) +
e^{{\cal V}^- (\F_0)}\right]\frac{\f_{(k)}^2 x^k}{\f_{(k)}^0 x^k}\,b^2 \,
e^{-\frac{1}{2}\left({\cal V}^+ (\F) +
{\cal V}^- (\F)\right)},\nonumber
\end{eqnarray}
\begin{eqnarray}
\bar{a}_{j}&=&\left[\sqrt{2}\left(
\frac{\f_{(k)}^1 x^k}{\f_{(k)}^0 x^k} e^{{\cal V}^+ (\F_0)} + 
\frac{\f_{(0)}^1}{\f_{(0)}^0}\right) - 
\frac{\lambda_1}{\sqrt{2}\lambda_3}\right]\f_{(j)}^0 b^1 \,
e^{-\frac{1}{2}(3{\cal V}^+ (\F) - {\cal V}^- (\F))}
\label{afncj}\\
&+&\left[\sqrt{2}\left(
\frac{\f_{(k)}^2 x^k}{\f_{(k)}^0 x^k} e^{-{\cal V}^- (\F_0)} +
\frac{\f_{(0)}^2}{\f_{(0)}^0}\right) -
\frac{\lambda_2}{\sqrt{2}\lambda_3}\right]\f_{(j)}^0 b^2 \,
e^{-\frac{1}{2}({\cal V}^+ (\F) - 3{\cal V}^- (\F))}
\nonumber\\
&+&\sqrt{2}\,\frac{\left(\f_{(j)}^1 \f_{(k)}^0 x^k - 
\f_{(j)}^0 \f_{(k)}^1 x^k\right)}{(\f_{(k)}^0 x^k)^2}\,
b^1 e^{{\cal V}^+ (\F_0)} \left(V^+ (\F) - V^+ (\F_0)\right)
e^{\frac{1}{2}({\cal V}^+ (\F) + {\cal V}^- (\F))}
\nonumber\\
&+&\sqrt{2}\,\frac{\left(\f_{(j)}^2 \f_{(k)}^0 x^k -
\f_{(j)}^0 \f_{(k)}^2 x^k\right)}{(\f_{(k)}^0 x^k)^2}\,
b^2 e^{-{\cal V}^- (\F_0)} \left(V^- (\F) - V^- (\F_0)\right)
e^{-\frac{1}{2}({\cal V}^+ (\F) + {\cal V}^- (\F))},\nonumber
\end{eqnarray}
and
\begin{eqnarray}
\Psi &=&\f_{(k)}^1 x^k \,e^{{\cal V}^+ (\F_0)}\,\left(
\lambda_1 -2\lambda_3 \frac{\f_{(0)}^1}{\f_{(0)}^0}\right)\,D^+ +
\f_{(k)}^2 x^k\,e^{-{\cal V}^{-}(\F_0)}\,\left(
\lambda_2 -2\lambda_3 \frac{\f_{(0)}^2}{\f_{(0)}^0}\right)\,D^- 
\nonumber\\
&-&\frac{\lambda_3}{\f_{(0)}^0}\,\left(\f_{(0)}^1 {\cal D}^+ +
\f_{(0)}^2 {\cal D}^- \right) + 
\lambda_3 \left({\cal C}^+ + {\cal C}^- \right) + 
\lambda_3 \f_{(k)}^3 x^k + \lambda_3 \f_{(0)}^3 x^0 +
\lambda_3 q^3 (0)\nonumber\\
&-&\frac{(\lambda_1)^2 \,V^+ (\F)
 + (\lambda_2)^2 \,V^- (\F)}{4\lambda_3}.\label{psifnc}
\end{eqnarray}
In the previous equations we set
\begin{eqnarray}
\phi = \f_{(0)}^0 x^0 + \f_{(k)}^0 x^k\qquad\qquad
\phi_0 = \f_{(0)}^0 x^0.
\end{eqnarray}
As for Eq.~(\ref{psifnc}), the following functions were 
introduced:
\begin{eqnarray}
{\cal C}^{+}&=&\frac{(\f_{(k)}^1 x^k)^2}{\f_{(k)}^0 x^k}\,
\left[1 - e^{2{\cal V}^+ (\F)}\,D^+ \right],\label{Cpiu}\\
{\cal C}^{-}&=&\frac{(\f_{(k)}^2 x^k)^2}{\f_{(k)}^0 x^k}\,
\left[1 - e^{-2{\cal V}^- (\F)}\,D^- \right],\label{Cmeno}\\
D^{\pm}&=&\frac{V^{\pm}(\F) - V^{\pm}(\F_0)}{\f_{(k)}^0 x^k},
\label{Dpm}\\
{\cal D}^{+}&=&\frac{\f_{(0)}^1}{\f_{(0)}^0}\,\left[
V^{+}(\F) - V^{+}(0)\right],\label{Dcalpiu}\\
{\cal D}^{-}&=&\frac{\f_{(0)}^2}{\f_{(0)}^0}\,\left[
V^{-}(\F) - V^{-}(0)\right].\label{Dcalmeno}
\end{eqnarray}

\section{Applications and Discussion}
\label{sec6}

In order to emphasize geometrical effects over cinematic ones we 
are naturally led to choose the coefficients $\f_{(\mu)}^\nu$ as 
follows:
\begin{eqnarray}
\f_{(0)}^{\mu}&=&-\delta^{\mu}_{\ 0} + \delta^{\mu}_{\ 3}\nonumber\\
\f_{(i)}^{j}&=&D^j_{\ i}\nonumber\\
\f_{(i)}^{0}&=&\f_{(i)}^{3}
\end{eqnarray} 
where
\begin{equation}
D^i_{\ k}\,D_{j}^{\ k} = \delta^i_{\ j}.
\label{orto}
\end{equation}
With this choice the observer is assumed to be at rest in the $y^\mu$ 
reference frame with given orientation with respect to $y^\mu$ in 
$q(\tau)$. Euler angles of rotation matrix $D^i_{\ k}$ determine the 
orientation. One can easily see that the conditions~(\ref{condfi}) are
met.

In order to obtain the relationship between Fourier coefficients 
$\bmit \lambda$ and the usual flat space--time wave--vector, we 
set $L=1$ and $\beta=0$ in formula~(\ref{psifnc}):
\begin{equation}
\Psi = \left(\lambda_j \,D^j_{\ k} - 
\frac{(\lambda_1)^2 + (\lambda_2)^2}{4\,\lambda_3}\,D^3_{\ k}
\right)\,x^k + \left(\lambda_3 + 
\frac{(\lambda_1)^2 + (\lambda_2)^2}{4\,\lambda_3}\right)\,x^0.
\end{equation}
In a flat space-time, $\Psi$ is usually defined as
\begin{equation}
\Psi = \Psi^{\pm} = k_j x^j \pm k x^0 \qquad\qquad k=
\sqrt{k_1^{\,2}+k_2^{\,2}+k_3^{\,2}}
\end{equation}
From the above equations, by taking into account the orthogonality of
matrix $D^i_{\ j}$ we obtain:
\begin{eqnarray}
\lambda_r&=&D_{r}^{\ j}\,k_j\nonumber\\
\lambda_3&=& \lambda_3^\pm = \frac{D_{3}^{\ j}\,k_j \pm k}{2}
\label{lambdak}
\end{eqnarray}
The integral of Eq.~(\ref{fourierfnc}) is expressed in terms of 
$\bmit k$ as a sum of two terms:
\begin{equation}
A_{\mu}=\int d^{3}k \,\left[e^{i\Psi^+}\,\bar{a}^+_{\mu} + 
e^{i\Psi^-}\,\bar{a}^-_{\mu}\right]
\end{equation}
where $\bar{a}^\pm_\mu=
\bar{a}_\mu\left[\bmit\lambda^\pm(\bmit k),x^\alpha; 
b_\pm^s(\bmit k)\right]$
with $\bar{a}_\mu$ given by Eqs.~(\ref{afnc0}) and~(\ref{afncj}). 

In order to better understand the behaviour of the solution, let us 
suppose:
\begin{equation}
D^{i}_{\ j} = \delta^i_{\ j}.
\end{equation}
In this case
\begin{equation}
\lambda_r=k_r\nonumber \qquad\qquad
\lambda^\pm_3=\frac{k_3 \pm k}{2},
\end{equation}
while
\begin{eqnarray}
\Psi^{\pm}&=&k_1 x^1 e^{{\cal V}^+ (-x^0)} D^{+} + 
k_2 x^2 e^{-{\cal V}^- (-x^0)} D^{-} +
\frac{k_3 \pm k}{2}\, \left(x^3 + x^0 \right)\nonumber\\ 
&-& \frac{(k_1)^2 V^+ (x^3 - x^0) + 
(k_2)^2 V^- (x^3 - x^0)}{2(k_3 \pm k)} +
\frac{k_3 \pm k}{2}\,\left({\cal C}^+ +
 {\cal C}^- \right),
\end{eqnarray}
where
\begin{eqnarray}
D^{\pm}&=&\frac{V^{\pm}(x^3-x^0) - V^{\pm}(-x^0)}{x^3}\nonumber\\
{\cal D}^{\pm} &=& 0 \nonumber\\
{\cal C}^+ &=&\frac{(x^1)^2}{x^3}\,\left[1 - 
e^{2{\cal V}^+ (x^3-x^0)}\,D^+\right]\nonumber\\
{\cal C}^- &=& \frac{(x^2)^2}{x^3}\,\left[1 -
e^{-2{\cal V}^- (x^3-x^0)}\,D^-\right]  
\end{eqnarray}

As an example we choose the constants $b_{\pm}^s$ so that there is a
static magnetic field along the $x^3$ direction in absence of 
gravitational wave. This is mathematically accomplished by setting
$b_{\pm}^r = b_{\pm}\,\delta^r_1$, where
\begin{equation}
b_+ = b_- = -\frac{i\,{\cal B}_0}{2\sqrt{2}\,k}\,
\delta(k_3 - k')\,\delta(k_1)\,\delta(k_2).
\end{equation}
By performing the integration and taking the limit 
$k' \rightarrow 0$, we achieve:
\begin{eqnarray}
A_0&=&-\frac{x^1 x^2}{x^3}\,{\cal B}_0 D^- \,e^{-{\cal V}^- (\mt)}\,
\left\{ e^{{\cal V}^+ (\mt)}\,e^{-\frac{1}{2}\,
[3{\cal V}^+ (\zt) - {\cal V}^- (\zt)]}\right.\nonumber\\
&+&\left.\left[\dot{\cal V}^+(\mt)\,e^{{\cal V}^+ (\mt)}\,\left(
V^+ (\zt) - V^+ (\mt)\right) - e^{-{\cal V}^+ (\mt)}\right]\,
e^{\frac{1}{2}\,[{\cal V}^+ (\zt) + {\cal V}^- (\zt)]}
\right\}\\
A_1&=&x^2 \,{\cal B}_0 \,D^+ D^- \,e^{\frac{1}{2}[
{\cal V}^+ (\zt) + {\cal V}^+ (\mt)]}\,
e^{\frac{1}{2}[{\cal V}^- (\zt) - {\cal V}^- (\mt)]}\nonumber\\
A_2&=&0\\
A_3&=&\frac{x^1 x^2}{x^3}\,{\cal B}_0 \,D^- \,e^{{\cal V}^+ (\mt)}\,
e^{-{\cal V}^- (\mt)}\,
\left\{e^{-\frac{1}{2}[3{\cal V}^+ (\zt) - {\cal V}^- (\zt)]} -
D^+ \,e^{\frac{1}{2}[{\cal V}^+ (\zt) + {\cal V}^- (\zt)]}
\right\}.
\end{eqnarray}
The above expressions hold true everywhere. In order to show their 
good behaviour in the neighborhood of the origin we perform a power 
expansion in $x^3$ up to $|\bmit x|^2$ terms. We obtain:
\begin{eqnarray}
A_0&=&x^1 x^2\,{\cal B}_0\,\dot{\cal V}^+ (\mt)\,
e^{-\frac{1}{2}\,[{\cal V}^+ (\mt) + {\cal V}^- (\mt)]}\label{a0xq}\\
A_1&=&x^2\,{\cal B}_0\,\left[1 - \frac{x^3}{2}\,\dot{\cal V}^+ (\mt) + 
\frac{3}{2}\,x^3\,\dot{\cal V}^- (\mt)\right]\,
e^{-[{\cal V}^+ (\mt) - 2\,{\cal V}^- (\mt)]}\\
A_2&=&0\\
A_3&=&-x^1 x^2\,{\cal B}_0\,\dot{\cal V}^+ (\mt)\,
e^{-\frac{1}{2}\,[{\cal V}^+ (\mt) -3\,{\cal V}^- (\mt)]}
\label{a3xq}
\end{eqnarray}

The electromagnetic tensor field components can be obtained from the 
usual relation:
\begin{equation}
F_{\mu\nu} = A_{\nu,\mu} - A_{\mu,\nu}
\end{equation}
Although it would be possible to calculate $F_{\mu\nu}$ everywhere, we 
shall limit ourselves near the origin.
Therefore, up to linear terms in $x^k$, we achieve:  
\begin{eqnarray}
F_{01}&=&x^2\,{\cal B}_0\,\left[\dot{\cal V}^+(\mt) - 2\,
\dot{\cal V}^-(\mt)\right]\,e^{-[{\cal V}^+(\mt)-2\,
{\cal V}^-(\mt)]}\nonumber\\ 
&-&x^2\,{\cal B}_0\,\dot{\cal V}^+(\mt)\,e^{-\frac{1}{2}\,
[{\cal V}^+(\mt) + {\cal V}^-(\mt)]}\label{f01}\\
F_{02}&=&-x^1\,{\cal B}_0\,\dot{\cal V}^+(\mt)\,
e^{-\frac{1}{2}\,[{\cal V}^+(\mt)+{\cal V}^-(\mt)]}\\
F_{03}&=&0\\
F_{12}&=&-{\cal B}_0\,\left[1-\frac{x^3}{2}\,\dot{\cal V}^+(\mt) +
\frac{3}{2}\,x^3\,\dot{\cal V}^-(\mt)\right]\,
e^{-[{\cal V}^+(\mt)-2\,{\cal V}^-(\mt)]}\\
F_{23}&=&-x^1\,{\cal B}_0\,\dot{\cal V}^+(\mt)\,
e^{-\frac{1}{2}\,[{\cal V}^+(\mt)-3\,{\cal V}^-(\mt)]}\\
F_{31}&=&x^2\,{\cal B}_0\,\left[-\frac{1}{2}\,\dot{\cal V}^+(\mt) +
\frac{3}{2}\,\dot{\cal V}^-(\mt)\right]\,e^{-[{\cal V}^+(\mt)-2\,
{\cal V}^-(\mt)]}\nonumber\\
&+&x^2\,{\cal B}_0\,\dot{\cal V}^+(\mt)\,e^{-\frac{1}{2}\,
[{\cal V}^+(\mt) - 3\,{\cal V}^-(\mt)]} \label{f31}
\end{eqnarray} 
We notice that in the laboratory frame the effect of a gravitational 
wave on an electromagnetic field is not only of tidal nature. In fact 
when $x^k\rightarrow 0$ the the only non--vanishing component becomes:
\begin{equation}
F_{12} = - B(x^0) = - {\cal B}_0 
e^{-[{\cal V}^+(\mt)-2\,{\cal V}^-(\mt)]}
\label{noneq}
\end{equation}
Two important conclusions may be drawn from this rather simple case.
First, the solution can not be interpreted as a photon creation.
In fact, the solution described by 
\mbox{Eqs. (\ref{f01})--(\ref{f31})} 
has non--vanishing d'Alembertian, proportional to the field itself. 
Moreover direct computation  
of the divergence of Pointing vector at the origin reads:
\begin{equation}
T^{0k}_{\ \ ,k} = - \frac{d\ \,}{dx^0} \left\{
\frac{{\cal B}_0^2}{8\,\pi}\,e^{-2\left[{\cal V}^+(-x^0) -
2\,{\cal V}^-(-x^0)\right]}\right\}\label{poin}
\end{equation}
It is the time derivative of a function assuming finite values. Its 
time average vanishes: in fact,
if the gravitational wave is a periodic function, the time average of 
Eq.~(\ref{poin}) over the period $T$ is zero; for a generic 
non--periodic wave, the time average vanishes over an interval of time 
that is longer than the signal duration 
(see for instance \cite{l2},
\S 34). Therefore there is no net flux of electromagnetic radiation 
across any close surface. This implies the absence of any radiation 
field. This conclusion is not surprising since
the solution still keeps the tensor form of a static magnetic 
field.

This is a good example in which the calculation in the framework of
the full theory of general relativity settles down an open question
of the linearized theory.
In fact, by using the linear approximation, the time average of the 
divergence of the the Pointing vector at the origin vanishes 
(taking only first order terms).
The first non--zero contribution is due to second order terms, 
which, however, are to be neglected in a linearized theory.
One could be led to think that quadratic terms could possibly give 
rise to a non--vanishing time average of $T^{0k}_{\ \ ,k}$. 
In the full theory, Eq. (6.27) immediately shows that this is not the 
case.

Another interesting feature is the possibility for an observer in FNC 
to prove --- at least in principle --- the presence of a gravitational 
wave, by performing an experiment involving electromagnetic 
interaction. The non--tidal nature of the interaction allows one to 
asses the presence of a gravitational wave irrespective of the 
apparatus size. This is quite 
different to mechanical detectors, involving 
geodesic deviation measurements. In the following we will propose a 
few {\em gedanken experiments} aimed at showing this peculiarity.

\subsection{Motion of a particle on the $x^3=0$ plane}

As a first example we consider a charged particle moving on the 
$x^3=0$ plane. The exact equations of motion are given by~\cite{l2}: 
\begin{equation}
mc\,\left(\frac{du^{\mu}}{ds} +\Gamma^{\mu}_{\alpha\beta}\,
u^{\alpha}u^{\beta}\right) = \frac{e}{c}\,F^{\mu}_{\ \alpha}\,
u^{\alpha}.
\end{equation}
As $\Gamma^{\mu}_{\alpha\beta}={\cal O}(x^k)$, we can 
assume that near the origin the leading term is the non--tidal one
[see Eq.~(\ref{noneq})].
Therefore equations of motion become ($\gamma = (1-\beta^2)^{-\frac{1}{2}}$, 
$\beta = \frac{v}{c}$, $v$ is the speed of the particle):
\begin{eqnarray}
\frac{du^0}{ds}&=&0\quad \Longrightarrow \quad u^0=\gamma 
\quad \Longrightarrow \quad x^0 = \gamma \,s,\\
\frac{du^3}{ds}&=&0\quad \Longrightarrow \quad u^3= 0 
\quad \Longrightarrow \quad x^3 = 0,\\
\frac{du^1}{dx^0}&=&-\frac{e}{{\cal E}}\,B(x^0)\,u^2,\\
\frac{du^2}{dx^0}&=&\frac{e}{{\cal E}}\,B(x^0)\,u^1,
\end{eqnarray}
where ${\cal E} = \gamma\,m\,c^2$ is the energy of the particle.
We see that the speed of the particle is a constant of
motion. Setting $B\,dx^0 = dw$ we get:
\begin{eqnarray}
u^1&=&\gamma \beta \,c_1\,\cos{\varphi (x^0)}
+ \gamma \beta \,c_2\,\sin{\varphi (x^0)},\\
u^2&=&\gamma \beta \,c_1\,\sin{\varphi (x^0)}
- \gamma \beta \,c_2\,\cos{\varphi (x^0)},\\
\varphi (x^0)&=&\frac{e}{{\cal E}}\,\left[
\int_0^{x^0}\,{B(\xi)d\xi} + w_0\right]\qquad\qquad
c_1^{\ 2} + c_2^{\ 2} = 1
\end{eqnarray}
the particle performs a non closed orbit around 
the origin, with a variable period $T(x^0)$ given implicitly by:
\begin{equation}
2\pi = \int_{x^0}^{x^0 + cT(x^0)}{\varphi (\xi)\,d\xi}
\end{equation}
By measurement of the time variation of the 
period (which is constant in flat space--time) one can infer the presence 
of a gravitational wave.

\subsection{Induced e.m.f. in conducting circuits}

As a second example we consider a conducting ring with 
resistance $R$, lying in the $x^3=0$ plane with its center at the 
origin (in this application SI units are used). Assuming the ring 
sufficiently small, we can neglect all the 
terms in the electromagnetic tensor except the one given by 
Eq.~(\ref{noneq}). Therefore Faraday law causes a current to flow. One 
has:
\begin{equation}
I(t) = -\,\frac{S}{R}\,\frac{\partial B(t)}{\partial t}
\end{equation}
where $S$ is the ring surface and $x^0 = c\,t$.

In general one expects the same kind of effect in RLC circuits. Let us 
consider a series circuit with a resistance $R$, an inductance 
${\cal L}$, and 
a capacitance $C$ under the same assumption as before. In this case 
the equation of motion for the charge on the condenser plates is:
\begin{equation}
\ddot{Q} + \frac{1}{\tau_0}\,\dot{Q} + 
\omega_0^{\ 2}\,Q = -\,\frac{S}{{\cal L}}\,\dot{B}
\label{RLC}
\end{equation}
where $\tau_0 = \frac{R}{{\cal L}}$,  
$\omega_0^{\ 2} = \frac{1}{{\cal L}C}$, and in this case the dot means 
derivation with respect to $t$. We 
notice that in flat space--time, where $\dot{B} = 0$, there is no 
current flowing. The effect of the gravitational wave is a 
current flowing through the circuit. 
In general the solution to Eq.~(\ref{RLC}) is given by:
\begin{equation}
{\cal Q}(t)=-\frac{S}{\sqrt{2\pi}}\,\int_{-\infty}^{+\infty}\,
e^{i\omega t}\,
\frac{i\omega\,B({\omega})}{\omega_0^2 - \omega^2 + 
i\,\frac{\omega}{\tau_0}} 
\label{sgen}
\end{equation}
($f(t) = \frac{1}{\sqrt{2\,\pi}}\int_{-\infty}^{+\infty}{e^{i\omega t}
f(\omega)\,d\omega}$). 

In order to study the main characteristics of the current we assume 
the gravitational wave to have a typical frequency $\omega_g$. In this 
way the Fourier components of the magnetic field are approximately 
given by:
\begin{equation}
B(\omega) \simeq \sqrt{2\pi}\,{\cal B}_0\,\left\{\delta(\omega) - 
\frac{i}{2}f_0\,\left[\delta(\omega-\omega_g) - \delta(\omega+\omega_g)
\right]\right\}
\label{Bomega}
\end{equation}
Substitution in Eq.~(\ref{RLC}) gives for the 
flowing current:
\begin{equation}
I(t) \simeq \frac{S\,{\cal B}_0\,\omega_g^{\ 2}\,f_0}{{\cal L}}\,
\Im{\left[\frac{e^{i\omega_g\,t}}{\omega_0^{\ 2}-\omega_g^{\ 2} + 
i\,\frac{\omega_g}{\tau_0}}\right]}
\stackrel{\omega_0=\omega_g}=-\frac{2\,S\,{\cal B}_0}{\cal L}\,
f_0\,{\cal Q}\,\cos{\omega_0\,t} 
\end{equation}
where ${\cal Q} = \frac{\omega_0\,\tau_0}{2}$ is the quality factor of 
the circuit.
Therefore it is possible, at least in principle, to detect a 
gravitational wave by measuring a current. It is noticeable that
the effect is not affected by the size of the device, but only by the
circuit parameters as it was already known in the framework of linear 
approximation (e.g.~\cite{mont96,mon98}).

\subsection{Paschen Back effect}

It is well known that the spectrum of Alkali atoms embedded in a 
static and homogeneous magnetic field shows frequency shifts. 
We are here interested in magnetic fields strong enough to let us 
neglect spin--orbit terms. Besides we assume that the system is so 
close to the origin that we do not have to take into account tidal 
terms (i.e. we can use the usual quantum mechanics in flat 
space--time). Therefore under these assumptions we start 
from the Hamiltonian describing the interaction with an external 
electromagnetic field in the non relativistic regime 
(e.g.~\cite{sakurai}). Substituting the value of the four--vector 
potential $A_\mu$ given in Eqs.~(\ref{a0xq})--(\ref{a3xq}), and neglecting
tidal effects we get:
\begin{eqnarray}
\hat{\cal H}&=&\hat{\cal H}_0 + \delta \,\hat{\cal H}\\
\hat{\cal H}_0&=&\frac{\hat{\bmit{P}}^2}{2m} + V(r)\\
\delta \,\hat{\cal H}&=&-\,\frac{e}{2\,mc}\,B(t)\,\left(
\hat{L}_z + 2\,\hat{S}_z\right)
\end{eqnarray}
where $e$ is the electron charge, $m$ the electron mass, and $V(r)$
the atomic potential energy. 
One could solve the problem using the standard time--dependent 
perturbation theory. However for typical frequencies of the 
gravitational wave much smaller than orbital ones (condition that 
is expected to hold true in nearly all circumstances), the effect 
results in a time variation of the frequency shifts of the flat
space--time spectrum. Thus we get:
\begin{equation}
\nu_{nl\rightarrow n'l'}(t)=\nu^{(0)}_{nl\rightarrow n'l'} - 
\frac{\mu_B\,B(t)}{\hbar}\,\Delta m_l\,;\qquad
\mu_B = \frac{|e|\hbar}{2mc}
\end{equation}
Under this assumption gravitational wave causes 
the distance between two lines of 
the usual Paschen Back spectrum to change with time.

\subsection{Spin precession}

Let us consider a free particle of charge $q$ and spin $\frac{1}{2}$. 
Neglecting all its degrees of freedom except for the spin, the interaction 
with an external magnetic field can be described introducing the 
following Hamiltonian:
\begin{equation}
\hat{\cal H} = -\frac{e}{m\,c}\,\bmit{B}(t)\cdot \hat{\bmit{S}} =
-\frac{e}{m\,c}\,B(t)\,\hat{S}_z
\end{equation}
Once again we neglect tidal effects. This is true provided the 
particle is sufficiently near the origin. Therefore the magnetic field 
is given in Eq.~(\ref{noneq}).
We obtain the evolution operator by solving Schr\"odiger 
equation. One has:
\begin{equation}
i\,\hbar\,\frac{\partial\ }{\partial t} \hat{\cal U} = 
\hat{\cal H}\,\hat{\cal U} \quad \Longrightarrow \quad
\hat{\cal U} = e^{-i\,\frac{2}{\hbar}\,\phi(t)\,\hat{S}_z},
\end{equation}
where
\begin{equation} 
\phi(t) = \frac{e\,{\cal B}_0}{2\,m\,c}\,\left[
t + F(t)\right];\qquad
F(t) = \int_0^t{dt'\,\left[\frac{B(t)}{{\cal B}_0} -1 \right]}
\end{equation}
Let us suppose to prepare the particle in the following state (at time
$t=0$):
\begin{equation}
|\alpha> = c_1\,|+> + c_2\,|->
\end{equation}
where $|\pm>$ are the eigenkets of the spin operator along $x^3$ axis.
Thus the state of the particle at time $t$ is given by:
\begin{equation}
|\alpha\,t> = \hat{\cal U}(t)\,|\alpha> = 
c_1\,e^{-i\,\phi(t)}\,|+> + c_2\,e^{i\,\phi(t)}\,|->
\end{equation}
Let
\begin{equation}
|x\,+> = \frac{1}{\sqrt{2}}\,\left(|+> + |->\right)
\end{equation}
be the eigenket of the spin operator along $x^1$ axis with positive 
eigenvalue~\cite{sakurai}. As it is well known, the probability to 
find the particle in such a state at time $t$ is given by
\begin{equation}
p(t) = \left|<x\,+|\alpha\,t>\right|^2 
= \frac{1}{2}\,\left[1 + 2\,|c_1|\,|c_2|\,
\cos{\left(\arg{c_2}-\arg{c_1}+2\,\phi (t)\right)}\right]
\end{equation}
If the magnetic field were static (i.e. in flat space--time) this 
probability would periodically vanish when ($n$ is integer):
\begin{equation}
t = T^{(0)}_n = \frac{(\arg{c_1}-\arg{c_2})\,mc}{e {\cal B}_0} +
\frac{mc}{e {\cal B}_0}
\,\arccos{\left(-\frac{1}{2\,|c_1|\,|c_2|}\right)} +
\frac{2\,mc\,\pi n}{e {\cal B}_0}
\end{equation}
However, because of the presence of a gravitational wave this 
probability does not vanish in $T^{(0)}_n$ any longer, but takes
values depending on $n$. Namely:
\begin{equation}
p(T^{(0)}_n) = \frac{1}{2}\,\left[1 - 
\cos{\left(\frac{e\,{\cal B}_0}{m\,c}\,F(T^{(0)}_n)\right)} - 
\sqrt{4\,|c_1|^2\,|c_2|^2 - 1}\,
\sin{\left(\frac{e\,{\cal B}_0}{m\,c}\,F(T^{(0)}_n)\right)}
\right]
\end{equation}

\section{Conclusions}

In this paper we have presented some interesting features of the 
behaviour of electromagnetic field in exact gravitational wave 
background by considering de Rham equations with no approximation.
We have expressed our results in the laboratory frame (FNC) where any 
measurable quantity should be referred to.

We have investigated a particular case in order to better understand 
the main features of the solution. In particular we have shown the 
appearance 
of non--tidal effects. Furthermore we have conceived some experimental
setups which--in principle--could outline the differences between the 
response of an electromagnetic device and a mechanical one. In fact 
the latter is related to the geodesic deviation, while the former 
includes spatial point--like effects as well. 
This behaviour is not in contrast with the principle of 
equivalence, which applies to arbitrarily small region of 
space--time.

It is also explicitly shown that in FNC the interaction does not 
create any photons. This is due to the fact that the d'Alembertian 
of the electromagnetic 
field does not vanish, being proportional to the field itself.


\acknowledgments

The authors are pleased to thank V. Guidi for reading of the 
manuscript. One of the author (E. M.) wish also to thank D. Etro for 
invaluable help.

\end{document}